\documentstyle[12pt,a4,epsf]{article}
\newcommand{\sig}{S}

\newcommand{\dps}{\displaystyle}
\newcommand{\tr}{\mbox{Tr}}
\newcommand{\sigg}{S^{\prime}}

\newcommand{\ra}{\rightarrow}
\newcommand{\si}{S_i}
\newcommand{\sj}{S_j}

\newcommand{\la}{\lambda}
\newcommand{\laa}{\lambda^\prime}
\newcommand{\zb}{\overline{z}}

\begin{document} 
\bibliographystyle{plain} 
\input{epsf} 
{\Large\bf\noindent Metastable states of a ferromagnet on random thin graphs} 
\vskip 2 truecm
\noindent A. Lefevre and D. S. Dean
\vskip 1 truecm
\noindent 
IRSAMC, Laboratoire de Physique Quantique, Universit\'e Paul Sabatier, 
 118 route de Narbonne, 31062 Toulouse Cedex 4.
\pagestyle{empty} 
\vskip 1 truecm \noindent{\bf Abstract:} 

We calculate the mean number of metastable states of an Ising ferromagnet on
random thin graphs of fixed connectivity $c$. We find, as for mean field spin
glasses that this mean increases exponentially with the number of sites,
and is the same as that calculated for the $\pm J$ spin glass on the same
graphs. An annealed calculation of the number $\langle N_{MS}(E)\rangle $ of
metastable states of energy $E$ is carried out. For small $c$, an analytic
result is obtained. The result is compared with the one obtained for 
spin glasses in order to discuss the role played by loops on thin 
graphs and hence the effect of real frustration on the distribution of 
metastable states. 

\vskip 1 truecm
\noindent
{\bf PACS:} 05.20 y Classical statistical mechanics, 75.10 Nr Spin glasses and
other random models. 
\vskip 1 truecm
\noindent  November  2000
\newpage
\pagenumbering{arabic} 
\pagestyle{plain}

\section{Introduction}

The nature of the spin glass phase is still a subject which is widely 
debated. Two possible scenarios have been proposed, one inspired from
mean field models which shows that in the spin glass phase one has
an extensive number of pure states, this phenomenon appears as 
replica symmetry breaking (RSB) in mean field models \cite{mepavi,fihe}. 
The other scenario is the droplet picture where there is a non extensive
number of pure states as in a ferromagnet \cite{fihu, bmscaling}.
If the RSB image is correct then an extensive number of 
pure states must also show up in the number of metastable states in 
a system, thought the relevance of metastable states or inherent states,
as they are referred to in the literature on glassy systems, to pure states 
is not obvious or even justified \cite{nest,monms}.  There has been a 
considerable amount of effort to analyze the metastable states in 
 mean field model 
\cite{taed,new,rob,dean,olfo,olfost}
 and also the number of solutions of the TAP mean 
field equations for this model (the generalization of metastable 
states to finite temperature) \cite{brmo1,brmo2}. 
Calculations on the Sherrington Kirkpartick (SK) totally connected spin glass
demonstrate the 
existence of an exponentially large (in terms of the number of spins $N$)
number of metastable states and the continuing existence of a macroscopic
entropy of metastable states even at arbitrarily high vales of a uniform 
magnetic field (in agreement with the divergence of the Almeida Thouless
line at zero temperature) \cite{alth}. This latter fact is clearly a pathology of the totally connected geometry of the SK model. In the SK model each spin 
is connected to all the other spins and the existence of the thermodynamic limit is ensured by scaling the couplings by a factor $1/\sqrt{N}$ in the case of
symmetric distributions. This scaling of the interaction strength with the 
system size is clearly undesirable when one wishes to make a connection with 
the finite dimensional analogue. Corrections to order
$1/c$, where c is the lattice connectivity, 
about mean field theory \cite{taed} seem to suggest an enhancement of the
number of metastable states when the dimension is reduced. Analytic 
studies of finite dimensional spin glasses are extremely difficult given that
the complexity of the starting point of any perturbative analysis, that is to 
say the Parisi replica symmetry breaking scheme. 

Recently there has been renewed interest in spin glasses on random 
graphs of finite
connectivity \cite{mon, mezpa, jon1,jon2, bcp}, 
the advantage with such systems is that while a mean field 
analysis is still possible, these systems mimic the finite connectivity of 
real finite dimensional spin glasses. It has been shown that the replica 
symmetric solution in such systems is not stable \cite{thou}.
Unfortunately no
exact treatment of the RSB solution has been achieved, there are however
approximate treatments which yield promising results \cite{mezpa,monrsb}.
 Additionally one may carry out a perturbative
replica symmetry breaking in some cases, such as close to the critical 
temperature or in the limit of large connectivity \cite{mot,dogo}. 
Interestingly it can 
be shown that the replica symmetric solution on thin graphs (random graphs
where each site has a fixed connectivity $c$) is equivalent to the solution
for a spin glass on a Cayley tree with branching ratio $c-1$
(see \cite{mezpa} and references therein), that is to say
the graph one would obtain roughly if one eliminated all the loops 
present in the corresponding random thin graph. It is well known that
the fraction of loops in such graphs goes as $\ln(N)$ where $N$ is the
number of sites. Therefore one can see, that despite the scarcity of 
such loops,
their effect is extremely important and that they make a replica symmetric 
system become RSB. Of course it is only through loops that one can have
real frustration \cite{toulouse}, without loops one may construct local
gauge transformations that make the system equivalent to a  ferromagnetic
one. 

In a recent paper \cite{dean2}, the number average metastable states of 
$\pm J$ (where each bond is taken to be $\pm J$ with probability $1/2$) 
spin glasses on random thin graphs has been calculated. At zero temperature
the number of metastable states is defined to be the number of spin
configurations stable to single spin flips. It was shown that this number
decreases as the connectivity is increased and in the limit $c\to \infty$
the result for the Sherrington Kirkpatrick mean field spin glass was 
recovered. In this paper we consider the problem of purely ferromagnetic
systems on such graphs. Here there is clearly no real frustration even with
loops. We find that the average total number of such metastable  states on the
ferromagnet is equal to  average total number on the corresponding
$\pm J$ spin glass. However one finds that when one calculates the average
number of metastable states of fixed energy $E$, $N_{MS}(E)$ there exists a
critical energy $E^*$ such that $\langle N^{SG}_{MS}(E)\rangle 
= \langle N^{F}_{MS}(E) \rangle$ for
$E\ge E^*$. (here the superscripts $F$ and $SG$ denote ferromagnet and 
spin glass respectively and $\langle \cdot \rangle$ denotes
the disorder average) but $\langle N^{SG}_{MS}(E)\rangle  
< \langle N^{F}_{MS}(E)\rangle$ for
$E < E^*$. Hence the rather surprising result that at lower energies
the ferromagnet has more metastable states than the spin glass. We show that
this difference is due to the effect of loops and moreover that $E^*$ is the
energy at which the metastable states of the ferromagnetic system 
acquire a non zero magnetisation. 
In addition we show that for $E>E^*$,  
$\ln\left(\langle N^{SG}_{MS}(E)\rangle\right)/N 
=\ln\left( \langle N^{F}_{MS}(E) \rangle\right)/N $ is a concave function of 
$E$ whilst for $E< E^*$, $\ln\left(\langle N^{SG}_{MS}(E)\rangle\right)/N$
remains concave but $\ln\left( \langle N^{F}_{MS}(E) \rangle\right)/N
$ becomes convex. 
Thus suggesting that the concavity of  $\ln\left( \langle N_{MS}(E) \rangle\right)/N$ at low energies and and hence temperature, may be an indication
of replica symmetry breaking.

\section{Analysis}

 The  model we shall consider has the Hamiltonian 
\begin{equation}
H = -{1\over 2} \sum_{j\neq i} J_{ij} n_{ij} S_i S_j
\end{equation} 
 where the $S_i$ are Ising spins, $n_{ij}$ is equal to one if the sites $i$ and $j$ are connected. In the spin glass case considered in \cite{dean}
the $J_{ij}$ are taken from a binary distribution where
$J_{ij}= -1$ with probability half and $  J_{ij}= 1$ with probability half.
In the ferromagnetic case we consider here one has $J_{ij} = 1$.
A metastable state is defined to be a configuration where if one changes the 
sign of any given spin the energy does not decrease, for the purposes of 
this paper we shall include the {\em marginal}, case where the energy does not 
change, as being metastable. With this definition number of metastable states is
given by \cite{taed,dean,rob}
\begin{equation}
 N_{MS}  = {\rm Tr} \prod_{i=1}^N \theta\left( \sum_{j\neq i} J_{ij}n_{ij} S_i S_j
\right)
\end{equation} 
The fact that we include the marginal case implies that here $\theta(x)$ the 
Heaviside step function is taken such that $\theta(0) =1$. In the spin glass
case one  may exploit
the parity of the distribution of the $J_{ij}$ by making a gauge transformation
$J_{ij} \to J_{ij}S_iS_j$  to obtain 
\begin{equation}
 \langle N_{MS} \rangle = 2^N\langle \prod_{i=1}^N \theta\left( \sum_{j\neq i} J_{ij} n_{ij}\right) \rangle
\end{equation}
However this is not possible in the ferromagnetic case, and thus renders
the ferromagnetic problem more difficult than the spin glass.

We shall use the method of  construction of the thin graphs
used in \cite{dean2}. Another method to 
generate these graphs by considering planar Feynman diagrams was used 
 in \cite{bcp,jon1,jon2}.
The random graphs are  constructed as follows: any two points are 
connected with probability $p/N$. Hence $n_{ij}$ is equal to one with probability $p/N$ and zero with probability $1-{p\over N}$. Here $p$ is some arbitrary number of order one and we shall see that the results one obtains are independent of the choice of $p$. If we denote the average on a random graph (with a specified value of $p$) by $\langle \cdot\rangle_p$ then the induced 
average over the subset of thin graphs of connectivity $c$ is given by 
\begin{equation} 
\langle F \rangle = 
{\langle  F \prod_{i=1}^N \delta_{\sum_{i\neq j} n_{ij} , c} \rangle_p\over
M(N,c,p)}
\end{equation}
where 
\begin{equation}
M(N,c,p) = \langle \prod_{i=1}^N \delta_{\sum_{i\neq j} n_{ij} , c} \rangle_p
\end{equation}
is the average number of thin graphs of connectivity $c$ generated by the 
random graph ensemble for a given $p$. Here, as opposed to the 
spin glass case, this is the only disorder average. It was shown in
\cite{dean2} that 
\begin{eqnarray*}
\ln(M(N,c,p)/N = \frac{c}{2}\,(\ln c- \ln p -1)-\ln(c!)-\frac{p}{2}
\end{eqnarray*}
With this averaging we therefore find that 
\begin{equation}
 \langle N^F_{MS} \rangle  = {D(N,c,p)\over M(N,c,p)}
\end{equation}
where 
\begin{equation}
D(N,c,p) =   \langle \prod_{i=1}^N \theta\left( \sum_{j\neq i} J_{ij} n_{ij}\right)   \delta_{\sum_{i\neq j} n_{ij} , c} \rangle_p  
\end{equation}
To compute $D(N,c,p)$ we introduce Fourier representations of the 
Heaviside and Kronnecker delta  functions to obtain:
\begin{eqnarray*}
D(N,c,p) &=&  \langle \int\,\frac{d\la_i\,dx_i\,dy_i}{(-4\pi^2)^N} {\rm Tr}_{\si} e^{
\sum_i\,\left( \la_i c +x_i y_j \right) 
-\frac{1}{2}\,\sum_{i \neq j}\,n_{ij}\,(\la_i+\la_j+y_i \si \sj)
} \rangle_p \\
	 &=& \int\,\frac{d\la_i\,dx_i\,dy_i}{(-4\pi^2)^N} {\rm Tr}_{\si} e^{
\sum_i\,\left( \la_i c
+x_i y_i \right) \prod_{i<j}\,\left[
1-\frac{p}{N}+\frac{p}{N}e^{-\la_i-\la_j-\frac{y_i+y_j}{2}\,\si\sj}
\right]} \\
	 &=& \int\,\frac{d\la_i\,dx_i\,dy_i}{(-4\pi^2)^N} {\rm Tr}_{\si}
e^{\sum_i\,\left(\la_i c+x_i y_i\right)}\,\exp\left[
-\frac{Np}{2}+\frac{p}{2N}\,\sum_{i \neq
j}\,e^{-\la_i-\la_j-\frac{y_i+y_j}{2}\,\si\sj}\right]
\end{eqnarray*}
as N goes to infinity. Here the integration ranges are $\lambda \in [0,2\pi]$,
$x\in [0,\infty]$ and $y\in [-\infty, \infty]$.
Now, we use  the useful identity: 

\begin{eqnarray}\label{identity}
e^{-\la-\laa-\frac{y+y^\prime}{2}\,\sig\sigg} &=& e^{
-\la-\laa}\,\left(
\cosh(\frac{y}{2})\cosh(\frac{y^\prime}{2})+\sinh(\frac{y^\prime}{2})\sinh(\frac{y}{2})
\right) 	\\ \nonumber
					      &+& e^{
-\la-\laa}\,\sig\sigg\left(
\cosh(\frac{y}{2})\sinh(\frac{y^\prime}{2})+\cosh(\frac{y^\prime}{2})\sinh(\frac{y}{2})
\right) 	
\end{eqnarray}
which allows us to write the term into brackets as
\begin{eqnarray*}
\mbox{exp}&& \left[\cdots \right] = \\ \nonumber 
\mbox{exp}&& \left[-\frac{Np}{2} +\frac{p}{2N}\,
{\left(\sum_i e^{-\la_i}\cosh(\frac{y_i}{2})\right)}^2+ \frac{p}{2N}\,
{\left(\sum_i e^{-\la_i}\sinh(\frac{y_i}{2})\right)}^2\right.\\ \nonumber
     &&+\left. \frac{p}{N}\,\left(\sum_i e^{-\la_i}\,\si\,\cosh(\frac{y_i}{2}) 
\right)\,\left(\sum_j \,\sj\,e^{-\la_j}\sinh(\frac{y_i}{2})\right)
\right]
\end{eqnarray*}
One can now decouple the sums by introducing two real Hubbard-Stratonovich 
fields $u$ and $v$ and a complex field $z$ giving

\begin{eqnarray*}
{D(N,c,p)} &=& \int\,dz\, d{\overline z}\,du\,dv\,e^{-N(\frac{u^2}{2}+\frac{v^2}{2}+|z|^2)}\\ \nonumber
         &\times&\left[
\frac{1}{-4\pi^2}\,{\rm Tr}_{\sig}\,\int\,d\la\,dx\,dy\,e^{\la c+y x
+\sqrt{p}\,e^{-\la}\,\left[(u+\sig\zb) \cosh(\frac{y}{2})+(v+\sig z) 
\sinh(\frac{y}{2})\right]}
\right]^N \nonumber
\end{eqnarray*}
where the trace above is over a single spin. 
By using the following identity:

\begin{equation}
\int\,\frac{d\la}{2i\pi}\,e^{\la c +\alpha e^{-\la}}=\frac{\alpha^c}{c!}
\end{equation}
we get:
\begin{eqnarray*}
&\,& \frac{1}{-4\pi^2}\,{\rm Tr}_{\sig}\,\int\,d\la\,dx\,dy\,e^{\la c+y x
+\sqrt{p}\,e^{-\la}\,\left[(u+\sig\zb) \cosh(\frac{y}{2})+(v+\sig z) 
\sinh(\frac{y}{2})\right]}\\ \nonumber
&=&\frac{1}{2i\pi\,c!}\,{\rm Tr}_{\sig}\,\int\,dx\,dy\,
{\left[\sqrt{p}\, (u+\sig\zb) \cosh(\frac{y}{2})+(v+\sig z) 
\sinh(\frac{y}{2})\right]}^c\\
&=&\frac{p^{\frac{c}{2}}}{2i\pi\,c!}\,\int\,dx\,dy\,\sum_{\sig\pm 1}
e^{y x}\,{\left[
\frac{e^{\frac{y}{2}}}{2}\,(u+\sig z+v+\sig\zb)+
\frac{e^{-\frac{y}{2}}}{2}\,(u+\sig z-v-\sig\zb)
\right]}^c
\end{eqnarray*}
Now, introducing 
$\dps{A=\frac{u+z+v+\zb}{2}}$, $\dps{B=\frac{u-z+v-\zb}{2}}$ and
$\dps{C=\frac{u+z-v-\zb}{2}}$, this 
term becomes 
\begin{eqnarray*}
\frac{p^{\frac{c}{2}}}{2c!}\,\left(C^c f(\frac{A}{C})+{\overline C}^c f(\frac{B}{\overline C})\right),
\end{eqnarray*}
where:
\begin{equation}
f(x)=\sum_{\frac{c}{2}\leq n \leq c} \left(
\begin{array}{rl}
c\\
n
\end{array}
\right)\,x^n.
\end{equation}
On calculating the ratio $D(N,c,p)/M(N,c,p)$ we find, as it should, that the
dependence on $p$ disappears and one obtains via a saddle point calculation 
in the large $N$ limit
\begin{equation}
\ln \left( \langle N^F_{MS} \rangle \right)/N = {\rm max}_{A,B,C} 
S^*(A,B,C)
\end{equation} 
where 
\begin{equation}
S^*(A,B,C)=-\frac{A^2+B^2}{2}-C{\overline C}+\ln\left(C^c f(\frac{A}{C})
+\overline C^c
f(\frac{B}{\overline C})\right) -{c\over 2}(\ln(c) - 1)
\end{equation}
We again change variables: $\dps{u=\frac{A}{C}}$, $\dps{v=\frac{B}
{\overline C}}$ and 
$\dps{t=\frac{\overline C}{C}}$, solving the saddle point equation
for $C$ and substituting in this solution yields 
\begin{equation}
\ln\left( \langle N^F_{MS} \rangle \right)/N = {\rm max}_{u,v,t} 
S_F(u,v,t)
\end{equation}
where 
\begin{equation}
S_F(u,v,t)=-\frac{c}{2}\,\ln\left(u^2+t^2 v^2+2t\right)
+\ln\left(f(u)+t^c\,f(v)\right).
\end{equation}
We notice $S_F(u,v,t)$ is invariant under the 
transformations $u \ra v$, $v \ra u$, $t \ra \frac{1}{t}$. There
is consequently a saddle point solution at the fixed point of this
transformation $u=v$ with $t=1$, this leads to exactly the saddle
point obtained for spin glasses where
\begin{equation}
S_{SG}(u)=-\frac{c}{2}\,\ln(1+u^2)+\ln(f(u))+(1-\frac{c}{2})\ln 2,
\end{equation}
In general one must solve the remaining saddle point equations 
numerically. For the case $c=1$ (dimers) and $c=2$, one dimensional 
chains the solution is identical to that for the spin glass
case \cite{dean2}. To continue we  will focus on the
$c=3$ case, which can also be computed analytically. In this case, the
stationarity conditions are:

\begin{eqnarray*}
u+2 &=& t\,(v+2)\\ \nonumber
t &=& \Psi (v)\\\nonumber
u^2 &=& t^3 v^2 \nonumber
\end{eqnarray*}
where $\dps{\Psi(v)=\frac{v+2}{v^2}}$. These equations imply the following:

\begin{eqnarray*}
U &\equiv& u+2\\ \nonumber
V &\equiv& v+2\\ \nonumber
U &=& \varphi (V)\\ \nonumber
V &=& \varphi (U), \nonumber
\end{eqnarray*}
where $\dps{\varphi(U)=\frac{U^2}{(U-2)^2}}$. This implies that $U$ and $V$ 
are solutions of $U=\varphi \circ \varphi (U)$, which has two kind of 
solutions:

\begin{itemize}
\item $U \neq V$~;
\item $U=V=\varphi (U)$, which is the one found for the spin-glass.
\end{itemize}
The first solution gives for the action the value $\dps{\frac{1}{2} \ln 
\frac{8}{7}}$, whereas the spin-glass one is $\frac{1}{2} \ln
(\frac{8}{5})$.
This shows that for $c=3$, the logarithm of the average number of metastable
states for the Ising model on random thin graphs is the same as the one
obtained for the $\pm 1$ spin-glass on the same graphs with an annealed
calculation. Carrying out a numerical investigation for $c > 3$ we find 
that $\langle N^F_{MS} \rangle =\langle N^{SG}_{MS} \rangle$. One may 
understand this result heuristically if one considers that $\langle N^F_{MS} \rangle$ and $ \langle N^{SG}_{MS} \rangle$ are dominated by the metastable
states at an energy where the  effect of loops is not important, then 
one may write
\begin{equation}
 \langle  N^F_{MS}\rangle  = \langle {\rm Tr_{S_i}} \prod_{i=1}^N \theta\left( \sum_{j\neq i} n_{ij} S_i S_j
\right)\rangle
\end{equation}
and
\begin{equation}
 \langle N^{SG}_{MS}\rangle   = \langle{\rm Tr}_{J_{ij}} \prod_{i=1}^N \theta\left( \sum_{j\neq i} n_{ij} J_{ij}
\right)\rangle
\end{equation}
where ${\rm Tr_{J_{ij}}}$ indicates a trace over independent dynamical 
variables $J_{ij}$ 
taking the values $\pm 1$ on the
bonds of the graph and the $J_{ij}$. In the ferromagnetic
case one may take the variables $S_i S_j$ to be independent variables 
taking the values $\pm 1$  {\em if one neglects the effects of loops which
would introduce correlations between these bond variables}. Hence
one expects that $\langle N^F_{MS} \rangle =\langle N^{SG}_{MS} \rangle $
if loops are not important at the energy level where the metastable states
are concentrated. In the  case $c=1$ and $c=2$ it is clear that loops 
cannot play a thermodynamically important role. One can make nonlocal gauge
transformations that in fact demonstrate that $ \ln (N^{SG}_{MS})/N
= \ln(N^F_{MS})/N$ with probability $1$ -- in the thermodynamic limit
 the two models are equivalent up to a gauge transformation.
We confirm this picture in the next section.

\section{Metastable states of fixed energy}

Here we define the average number of metastable states of fixed energy $NE$

\begin{equation}
 N_{MS}(E)  = {\rm Tr} \prod_{i=1}^N \theta\left( \sum_{j\neq i}
 J_{ij}n_{ij} S_i S_j\right) \delta(H - NE).
\end{equation} 

To achieve this, we need to introduce a Lagrange multiplier $\alpha$ to fix
the energy and carry out a calculation almost identical to that
of the previous section.

We find

\begin{equation} 
\frac{\ln\left( \langle N^F_{MS}(E) \rangle \right)}{N} = 
{\rm max}_{u,v,t,\alpha} S(u,v,t,\alpha;E)
\end{equation}
where 

\begin{eqnarray*}
S(u,v,t,\alpha;E)=-\frac{c}{2}\,\ln (u^2+t^2 v^2+2t)+\ln \left( 
f(ue^{-\alpha})+t^c \,f(ve^{-\alpha}) \right)+\alpha \, (\frac{c}{2}-E)
\end{eqnarray*}

\begin{figure}[htb]
\begin{center}\leavevmode
\epsfxsize=15 truecm\epsfbox{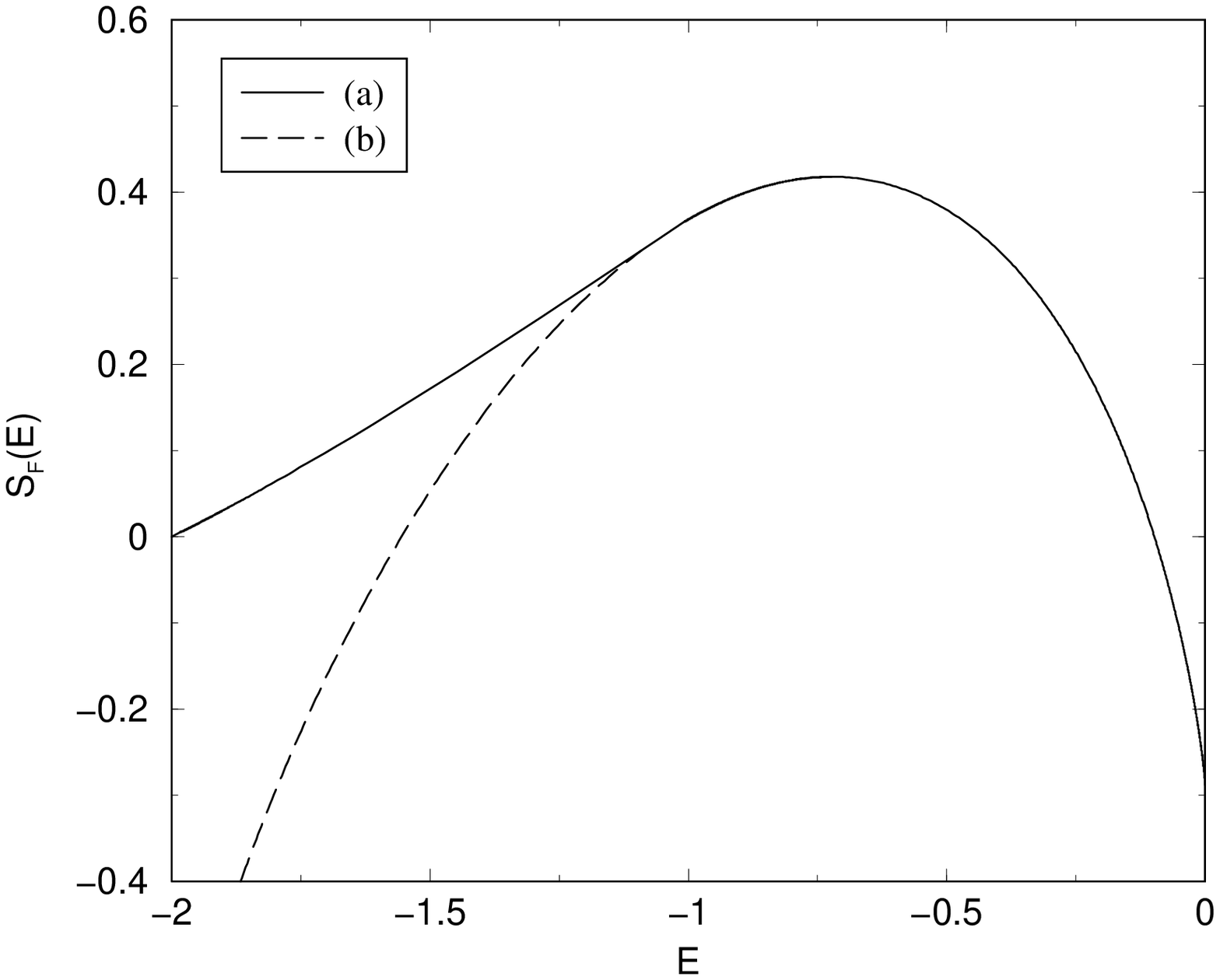}
\end{center}
\caption{Numerical calculation of $S_F(E) = 
\ln\left(\langle N_{MS}(E)\rangle\right)/N$ of the number of metastable 
states of fixed energy for $c=4$ (a). The dashed line (b) is the corresponding 
solution for the spin glass.} 
\label{c4nenising}
\end{figure}

At the saddle-point, the energy is :
\begin{eqnarray*}
E=\frac{c}{2}\frac{2t-u^2-t^2 v^2}{2t+u^2+t^2 v^2}.
\end{eqnarray*}
We call $x=\frac{2E}{c}$, so we have the relation:
\begin{eqnarray}\label{nrj}
\frac{u^2+t^2 v^2}{2t}=\frac{1-x}{1+x}.
\end{eqnarray}
For $c=2$, the result is the same as for the spin-glass  as expected. Let us
show briefly how to recover this with a transfer matrix method on the
one dimensional model. First,
one makes the  gauge transformation $\sig_i \sig_{i+1}\ra \sig_i$.
One introduces a Lagrange multiplier $\alpha$ to fix the energy and we get:
\begin{eqnarray*}
N_{MS}(E)={\rm max}_\alpha \,\tr\, {\cal M}^N_{\alpha},
\end{eqnarray*}
where 
\begin{eqnarray*}
{\cal M}_\alpha(\sig,\sigg)=e^{\alpha E+\alpha\frac{\sig+\sigg}{2}}\,
\theta(\sig+\sigg).
\end{eqnarray*}
So $\frac{\ln(N_{MS}(E))}{N}={\rm max}_\alpha \ln(\mu_\alpha)$, where 
$\mu_\alpha$ is the largest eigenvalue of ${\cal M}_\alpha$. We find:
\begin{eqnarray*} 
\mu_\alpha=\frac{e^\alpha+\sqrt{e^{2\alpha}+4}}{2}
\end{eqnarray*}
So at the maximum: $\alpha^*=\ln\frac{-2E}{\sqrt{1-E^2}}$ and we recover the
result of \cite{dean2} for the $c=2$ spin glass $\pm J$ on a thin graph
\begin{eqnarray*}
\frac{\ln(N_{MS}(E))}{N}=\frac{1-E}{2}\ln(\frac{1-E}{2})
-\frac{1+E}{2}\ln\frac{1+E}{2}+E\ln(-2E),
\end{eqnarray*}
confirming the above assertion.
For generic values of the local connectivity, the saddle-point equations can
be solved numerically. In fig.(\ref{c4nenising}), we have plotted the result
for $c=4$.
The curve corresponding to $c=3$ has also been calculated numerically
and agrees perfectly with the following calculation.
Let us focus onthe $c=3$ case.
For convenience, we introduce new variables:
\begin{eqnarray}
U &=& ue^{-\alpha}+2 \\ 
V &=& ve^{-\alpha}+2\\ 
a &=& \frac{e^{-2\alpha}}{2}  \label{eq:alpha}
\end{eqnarray}
In this case, the stationarity conditions lead to:

\begin{eqnarray}\label{newcol}
U &=& tV \\ \nonumber
t &=& \frac{ve^{-\alpha}+2}{v^2}=\frac{u^2}{ue^{-\alpha}+2}
\end{eqnarray}
and the function $\varphi$ is now $\dps{\frac{U^2\,e^{-2\alpha}}{{(U-2)}^2}}$.
The equation $\varphi \circ \varphi (U)=U$ is of degree four and can be
factorised by the second degree equation $\varphi (U)=U$. There are two
solutions with $U\neq V$ obeying

\begin{eqnarray}\label{uv}
\frac{U+V}{2} &=& \frac{2-a}{(1-a)^2}\\ \nonumber
UV &=& \frac{4}{(1-a)^2}.
\end{eqnarray}

\begin{figure}[htb]
\begin{center}\leavevmode
\epsfxsize=15 truecm\epsfbox{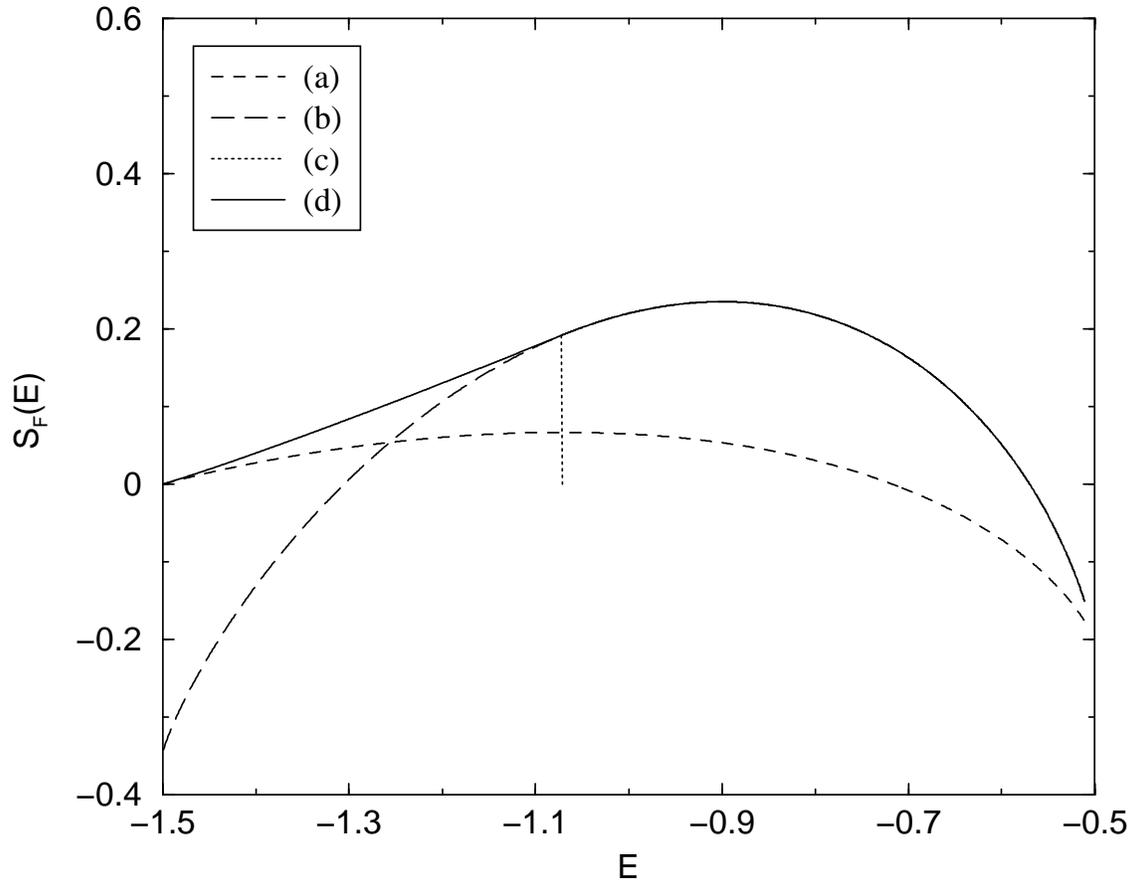}
\end{center}
\caption{$S_F(E) = 
\ln\left(\langle N_{MS} (E)\rangle\right)/N$ for $c=3$ shown by 
the solid curve (d).
Also plotted is the solution corresponding to $\alpha_-$ (a), 
the corresponding spin glass
solution in the low energy region (b). The vertical 
dotted line (c) represents the crossover point where the spin glass and 
ferromagnetic solutions start to differ. } 
\label{c3nenising}
\end{figure}

Moreover, by using (\ref{newcol}) in (\ref{nrj}) one obtains
\begin{eqnarray*}
\frac{u^2+t^2 v^2}{2t}=\frac{U+V}{2}=\frac{1-x}{1+x}
\end{eqnarray*}
yielding two different values for $a$:


\begin{eqnarray*}
a_{\pm} = \frac{1-3x \pm \sqrt{(1+x)(5-3x)}}{2(1-x)}, 
\end{eqnarray*}
yielding two possible solutions $\alpha_+$ and $\alpha_-$ from equation
(\ref{eq:alpha}).
In fig.(\ref{c3nenising}), we have plotted the value of the action obtained 
by solving the remaining 
stationary conditions for the two different values of $\alpha$ and the
annealed calculation for the $\pm 1$ spin-glass (there is only one 
real  solution for $U=V$).
The curve corresponding to $\alpha_-$ gives defined values for $x$ between -1
and $-\frac{1}{3}$, and reaches a maximum at $x^* = -5/7$ which value is 
$\dps{\frac{1}{2} \ln \frac{8}{7}}$ and corresponds to the value obtained
for $U\neq V$ in the previous calculation of the total complexity. This
solution however always has an action of lower value than that coming from
the spin glass solution. The solution coming from 
$\alpha_+$ is more pathological. 
Above the value  $x^*= -5/7$   the solution corresponding to
$\alpha_+$ does not exist
(this value of $x^*$ is the value over which $U$ and $V$
obtained from $\alpha_+$ become imaginary and so the corresponding value of
the action $S$ not real). The value of $E$ corresponding to this $x^*$ is 
shown by the vertical dotted line on fig.(\ref{c3nenising}). For $x < x^*$
the action corresponding to the solution with $\alpha_+$ is greater than
that coming from the spin glass saddle point and hence dominates in the 
thermodynamic limit. One should also note that this action becomes 
equal to zero at $E=-3/2$ the ground state for the ferromagnet, as it should.
Hence we see that the energy level with the largest number of
metastable states, and thus dominating the average total number, occurs
at any energy higher than that where the difference between the spin glass
and ferromagnetic calculations yields different results and above this energy
level the effect of loops is negligible. However, below $x^*$ the number of
metastable states is larger in the ferromagnet than in the spin glass.
In addition we shall  see that $E^*$ is the energy below
which the metastable states acquire a non-zero global magnetisation.

We now  continue the computation of the number of metastable states of
fixed energy $E$ but with fixed magnetisation $m = {1\over N}\sum S_i$.
The average number of  metastable states of energy $E$ and magnetisation
$M$ is then given by  
\begin{eqnarray*}
N_{MS}(E,m)=\tr \prod_{i=1}^N \theta\left( \sum_{j\neq i} J_{ij}n_{ij} S_i S_j
\right) \delta(H-NE) \delta (Nm-\sum_i \sig_i)
\end{eqnarray*}
One introduces another Lagrange multiplier $h$ to fix with $m$. The
resulting action is now

\begin{eqnarray*}
S^*_F(u,v,t,h,\alpha;E,m)&=&-\frac{c}{2}\ln(u^2+t^2 v^2+2 t)+
\ln(f(ue^{-\alpha})\, e^h+t^c\, f(ve^{-\alpha})\, e^{-h})\\ \nonumber
&+&\alpha(\frac{c}{2}-E)-mh.
\end{eqnarray*}
\begin{eqnarray*}
{\ln \left(N_{MS}(E,m)\right)\over N}={\rm max}_{u,v,t,h,\alpha} S^*_F(u,v,t,h,\alpha;E,m)
\end{eqnarray*}
The stationarity condition with respect to $h$ gives

\begin{eqnarray}
m=\frac{f(ue^{-\alpha})\, e^h- t^c\, f(ve^{-\alpha})\, e^{-h}}
{f(ue^{-\alpha})\, e^h+ t^c\, f(ve^{-\alpha})\, e^{-h}}
\end{eqnarray}	

substituting this values for $h$ in the action yields the reduced action
\begin{eqnarray*}
S^*_F(u,v,t,\alpha;E,m)&=&-\frac{c}{2}\ln(u^2+t^2 v^2+2 t)
+\frac{1+m}{2}\ln(f(ue^{-\alpha}))
+\frac{1-m}{2}\ln(f(ve^{-\alpha}))\\
&+& c\frac{1-m}{2}\ln t +\alpha(\frac{c}{2}-E)
-\frac{1+m}{2}\ln(\frac{1+m}{2})
-\frac{1-m}{2}\ln(\frac{1-m}{2}).
\end{eqnarray*}

The remaining stationarity conditions are 

\begin{eqnarray*}
U &=&\frac{2(m+1)}{1+x}-1 \\
V &=&\frac{2(1-m)}{1+x}-1 \\
\alpha &=&\frac{1}{4}\ln(\frac{UV}{(U-2)^2 (V-2)^2}) \\
t &=&\frac{u^2}{U}=\frac{V}{v^2} \\
\end{eqnarray*}

\begin{figure}[htb]
\begin{center}\leavevmode
\epsfxsize=15 truecm\epsfbox{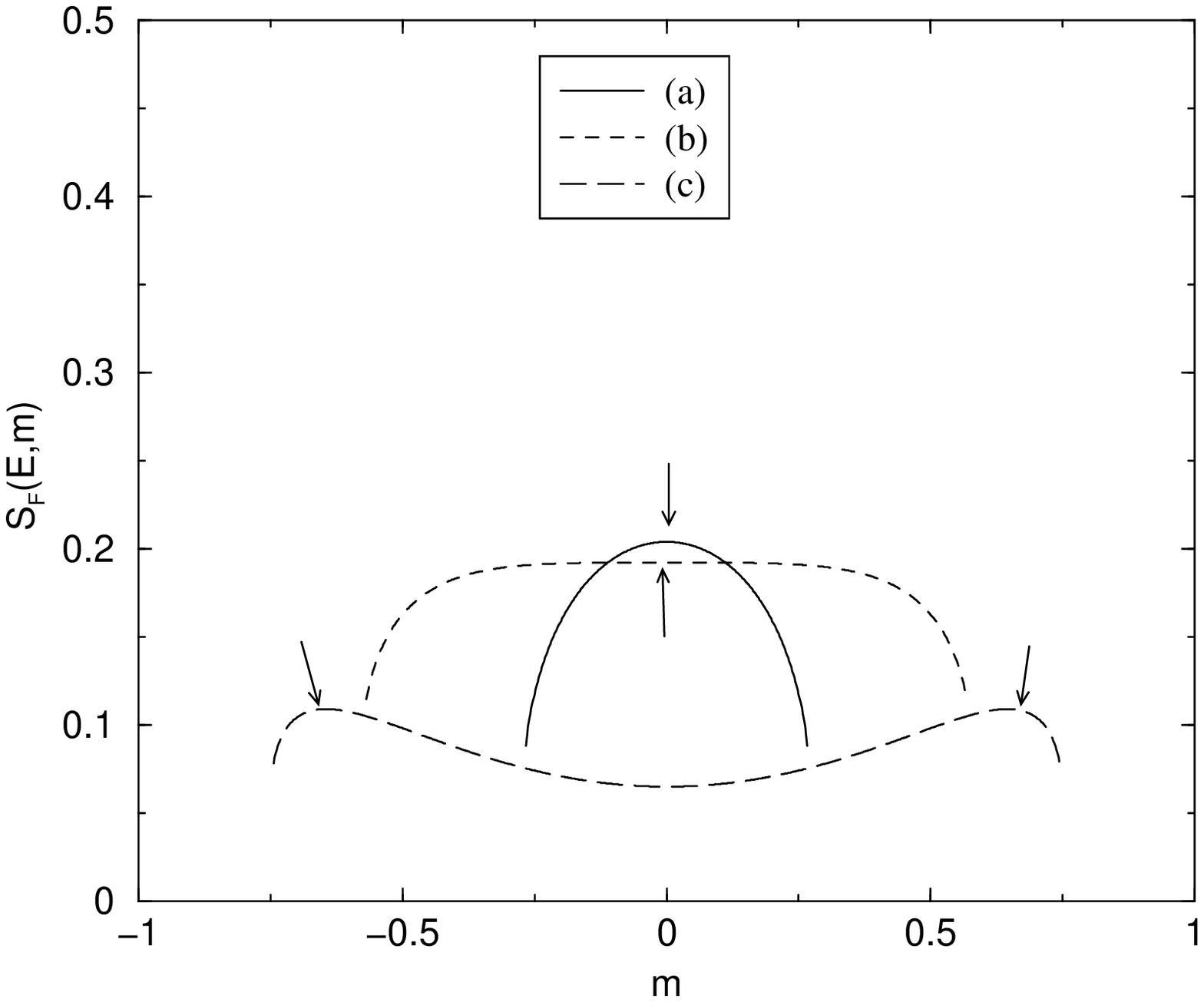}
\end{center}
\caption{ $S_F(E,m) = 
\ln\left(\langle N_{MS}(E,m)\rangle\right)/N$ for $c=3$ as a function
of the magnetisation $m$ for 
different values of the energy: $-0.765 > E^*$ (a), $-1.071 = E^*$ (b),
$-1.245 < E^*$ (c). The arrows indicate the local maximum which gives the
dominant contribution to $S_F(E)$ at fixed energy. At $E\simeq E^*$, the
maximum corresponding to the spin glass solution splits into two maxima and
the $Z_2$ symmetry is spontaneously broken.} 
\label{metaisingwithm}
\end{figure}

For a fixed energy greater than the ground state,
the number of metastable states must go to zero before  $m^2=1$. Indeed, when
$m+1=\frac{3}{2}(1+x)$ and $1-m=\frac{3}{2}(1+x)$, $u$ and $v$ are
 respectively zero, $S^*_F$
exhibits a singularity  hence values of $|m| > -\frac{1+3x}{2}$ are excluded.
Now, if we fix the energy and plot the number of metastable states for $m$
going from $-1$ to $1$, we get two kinds of configurations:

\begin{itemize}
\item if $x \geq x^*$, then the maximum is at $m=0$, that is the
magnetisation is zero~;
\item if $x \leq x^*$, then the point $m=0$ is a local minimum, and
there are two local maxima of opposite non zero magnetisations.
\end{itemize}
These results are demonstrated in the different regimes in fig. 
(\ref{metaisingwithm}).
The stationarity condition on $m$ leads to:
\begin{eqnarray*}
m=\frac{f(ue^{-\alpha})-t^c\,f(ve^{-\alpha})}
{f(ue^{-\alpha})+t^c\,f(ve^{-\alpha})},
\end{eqnarray*}  
which leads to a second order  transition in the value of $m$ at $x^*$.

One finds therefore, by comparison at the same energy $E$ with the spin glass, 
that the possibility of a non zero magnetisation paradoxically increases the
metastability of the system. 
\section{Conclusion}

In conclusion, we have seen that the mean number of metastable states for
the Ising ferromagnet on thin graphs increases exponentially with the size
of the system. Moreover, for the total average number, 
the result is the same as the one
obtained for the corresponding $\pm J$ spin glass. 
The complexity does change in the low
energy phase where it becomes convex in the case of the ferromagnet. This
shows that at high energy, the metastability  is mainly due to the 
local geometry of the graphs, and the relevance of  loops seems not to be 
significant.
At low energy, the presence of some non-zero magnetisation 
for $c>2$ seems to be
responsible of a complexity bigger than the one computed (in an annealed
calculation) for the spin glass.

\newpage
\pagestyle{empty}

\newpage
\pagestyle{empty}

\newpage
\pagestyle{empty}


\baselineskip =18pt

\begin{thebibliography}{0}
\bibitem{mepavi}{ M\'ezard M ,  Parisi G and  Virasoro M A 1987 {\em Spin Glass Theory and Beyond} (Singapore: World Scientific) }
\bibitem{fihe}{ Fischer K H and Hertz J A 1991 {\em Spin Glasses} (Cambridge
University Press)}
\bibitem{fihu}{ Fisher D S and Huse D A 1986 {\em Phys. Rev. Lett.} {\bf 56}
 1601}
\bibitem{bmscaling}{Bray A J and Moore M A 1987 {\em Phys. Rev. Lett.}
{\bf 58} 57}
\bibitem{nest}{Newman C M and Stein D L 1999 {\em cond-mat}, 9908455}
\bibitem{monms}{ Biroli G and Monasson R 2000 {\em Europhys. Lett.} {\bf 50} 
(2) 155}
\bibitem{taed}{Tanaka F and Edwards S F 1980 {\em J. Phys. F.} {\bf 13} 2769}
\bibitem{new}{de Dominicis C, Gabay M, Garel T and Orland H 1980 
\bibitem{rob}{Roberts S A 1981 {\em J. Phys. C.} {\bf 14} 3015}
\bibitem{dean}{ Dean D S 1994 {\em J. Phys. A.}  {\bf 27} L899}
{\em J. Physique} {\bf 41} 923}
\bibitem{olfo}{de Oliveira V M and Fontanari J F 1997 {\em J. Phys. A.} {\bf 30} 8445}
\bibitem{olfost}{de Oliveira V M,  Fontanari J F and Stadler P F 1999 {\em cond-mat 9908439}}
\bibitem{brmo1}{ Bray A J and Moore M A 1980 {\em J. Phys. C.} {\bf 13} L469}
\bibitem{brmo2}{Bray A J and Moore M A 1981 {\em J. Phys. C.} {\bf 14} 1313}
\bibitem{alth}{de Almeida J R L and Thouless D J 1978 {\em J. Phys. A.}
{\bf 11} 983}
\bibitem{mon}{Monasson R 1998 {\em J. Phys. A.} {\bf 31} 513}
\bibitem{mezpa}{ M\'ezard M and Parisi G 2000 {\em cond-mat/0009418}}
\bibitem{jon1}{Janke W and Johnston D 2000
{\em Nucl.Phys. B} 578  681-698}
\bibitem{jon2}{ Baillie C F, Janke W Johnston D.A and Plechac P
1995 {\em Nucl. Phys. B} 450  73}
\bibitem{bcp}{Bachas C de Calan C and Petropoulos P 1994 {\em J. Phys. A.}
{\bf 27} 6121}
\bibitem{thou} {Thouless D J 1986 {\em Phys. Rev. Lett.} {\bf 56} 1082}
\bibitem{monrsb}{Biroli G,  Monasson R and Weigt M 2000 
{\em Eur. Phys. J. B} {\bf 14}  551} 
\bibitem{mot}{ Mottishaw P 1988 {\em Europhys. Lett.} {\bf 3} 333}
\bibitem{dogo}{ De Dominicis C and  Goldschmidt Y Y 1989  {\em J. Phys. A.}
{\bf 22} 775}
\bibitem{toulouse}{Toulouse G 1977 {\em Commun. Phys.} {\bf 2} 115}
\bibitem{dean2}{Dean D S 2000 {\em Eur. Phys. J. B}{\bf 15} 493}
\bibitem{vibr}{ Viana L and Bray A J 1985 {\em J. Phys. C.} {\bf 18} 3037}
\bibitem{instab}{Mottishaw P and de Dominicis C 1987 {\em J. Phys. A}
{\bf 20} L375}
\end{thebibliography}
\end{document}